\documentclass[12pt,preprint]{aastex}

\slugcomment{This article is accepted for publication in
\textbf{MNRAS}}

\shorttitle{Magnetic fileds in IRDCs}

\shortauthors{Vahdanian and Nejad-Asghar}

\begin{document}

\title{Some Aspects of Rotation and Magnetic Field Morphology in the Infrared Dark Cloud G34.43+00.24}

\author{Hamed Vahdanian, Mohsen Nejad-Asghar}

\affil{Department of Theoretical Physics, Faculty of Science,
University of Mazandaran, Babolsar, Iran}

\email{nejadasghar@umz.ac.ir}

\begin{abstract}
The infrared dark clouds (IRDCs) are molecular clouds with
relatively greater values in their magnetic field strengths. For
example, the IRDC G34.43+00.24 (G34) has magnetic field strength of
the order of a few hundred micro-Gauss. In this study, we
investigate if the dynamic motions of charged particles in an IRDC
such as G34 can produce this magnetic field strength inside it. The
observations show that the line-of-sight velocity of G34 has global
gradient. We assume that the measured global velocity gradient can
correspond to the cloud rotation. We attribute a large-scale current
density to this rotating cloud by considering a constant value for
the incompleteness of charge neutrality and the velocity differences
between the positive and negative particles with very low ionization
fractions. We use the numerical package FISHPACK to obtain the
magnetic field strength and its morphology on the plane-of-sky
within G34. The results show that the magnetic field strengths are
of the order of several hundred micro-Gauss, and its morphology in
the plane-of-sky is somewhat consistent with the observational
results. We also obtain the relationship between magnetic field
strength and density in G34. The results show that with increasing
density, the magnetic field strength increases approximately as a
power-law function. The amount of power is approximately equal to
$0.45$, which is suitable for molecular clouds with strong magnetic
fields. Therefore, we can conclude that the dynamical motion of
IRDCs, and especially their rotations, can amplify the magnetic
field strengths within them.
\end{abstract}


\section{Introduction}

The infrared dark clouds (IRDCs) were discovered three decades ago,
as silhouettes against a bright infrared background, by the Infrared
Space Observatory (Perault et al.~1996) and the Midcourse Space
Experiment (Egan et al.~1998). Carey et al.~(1998) were the first
group to study the physical properties of IRDCs after their
discovery. They investigated $10$ observed clouds, all $1$ to $8\,
\mathrm{kpc}$ away from us, with diameters ranging from $0.4$ to
$15\, \mathrm{pc}$.  Their analysis on the extinction of the
observed lines indicated that the IRDCs are cold and dense regions,
with temperatures less than $20\, \mathrm{K}$ and density larger
than $10^5 \, \mathrm{cm}^{-3}$. Rathborne et al.~(2006) scrutinized
the millimeter continuum maps of $38$ IRDCs in our galaxy. These
clouds range in morphology from filamentary to compact and have
masses of $120$ to $16000\, \mathrm{M}_\odot$. Within these IRDCs,
they also found $140$ cores, with masses in the range of $10-2100\,
\mathrm{M}_\odot$, which their mass spectrum slope was similar to
that of the stellar IMF. Therefore, they concluded that IRDCs could
be suitable precursors to form stellar clusters and massive stars.
The observations show that the IRDCs are magnetized molecular
clouds. By developing the observational techniques in the last
decade, more information about the magnetic field morphology in the
IRDCs has been obtained (e.g., Santos et al.~2016, Hoq et al.~2017,
Beuther et al.~2018, Juvela et al.~2018, Liu, Zhang, \& Qiu et
al.~2020, A\~{n}ez-L\'{o}pez et al.~2020).

One interesting IRDC is G34.43+00.24 (hereafter G34), which for the
first time had been probed by Rathborne et al.~(2005). This cloud,
which is $3.6\, \mathrm{kpc}$ away from us and is located on the
Galactic plane, has a filamentary clumpy structure (e.g., Liu,
Sanhueza \& Liu et al.~2020, Isequilla et al.~2021). Four major
clumps of the G34 are known as MM1-MM4, which are nurseries for star
formation with masses between $330$ and $1500\, \mathrm{M}_\odot$
and number densities between $1.8 \times 10^4$ and $3.9 \times 10^5
\, \mathrm{cm}^{-3}$ (Sanhueza et al.~2010). By developing
technology and receiving observational information on the smaller
scales of the IRDCs, Jones et al.~(2016) used FORCAST on SOFIA and
submillimeter polarimetry to probe inside the MM1 clump of the G34.
Their observations showed that G34 is embedded in an external
magnetic field parallel to the Galactic plane and that the magnetic
field inside MM1 is perpendicular to the long axis of G34.

Observational information, such as velocity, density, magnetic
field, etc., corresponds to the projected information of the 3D
medium. The projection is made along the line-of-sight generating a
2D plane-of-sky field. Tang et al.~(2019) used thermal dust
polarization at $350\mu \mathrm{m}$ with an angular resolution of
$10''$ ($0.18 \, \mathrm{pc}$) to observe the plane-of-sky magnetic
field morphology across G34. Their investigations showed that the
$B$-fields across the clumps MM1 and MM2 are mostly perpendicular to
the filament's axis, with a prevailing orientation around
$70^\circ$. The MM3, oriented with its longer axis about $40^\circ$
off the north-south axis, reveals a $B$-field that is changing its
orientation to close to parallel to the clump's dust ridge. Soam et
al.~(2019) also used $850 \,\mu \mathrm{m}$ polarized dust emission
at the James Clerk Maxwell telescope to present the $B$-fields 2D
mapped in G34. They obtained the plane-of-sky magnetic field
strengths of $470\pm190 \, \mu \mathrm{G}$, $100\pm40 \, \mu
\mathrm{G}$, and $60\pm34 \, \mu \mathrm{G}$ in the central,
northern, and southern regions of G34, respectively.

All the observational data show that the magnetic field strengths in
the IRDCs are of the order of several hundred $\mu \mathrm{G}$. On
the other hand, we know that all the interstellar clouds and also
IRDCs are immersed in the Galactic magnetic fields. The strengths of
the magnetic fields in the Galactic plane are of the order of
several $\mu \mathrm{G}$ (Jansson \& Farrar~2012). Now the question
is, what is the reason for the increase in magnetic field strength
from the order of a few $\mu \mathrm{G}$ outside an IRDC to a few
hundred $\mu \mathrm{G}$ inside? One way to answer this question is
to use the topic of flux freezing in the ideal MHD. In a way that if
an IRDC is formed from a diffuse ISM due to compression, with
assuming the flux freezing, an increase in density from $n_i$ to
$n_f$ can increase the magnetic field by $B_f \sim B_i
(n_f/n_i)^{2/3}$ (e.g., Mestel~1966). But the ionization fraction in
the molecular clouds (such as IRDCs) is low, and the non-ideal MHD
effects (such as ambipolar diffusion) are impressive (e.g.,
Hennebelle \& Inutsuka~2019). On the other hand, we know that an
IRDC has dynamic motions and is ionized by various mechanisms so
that it contains moving ions, electrons, and charged grains. So we
are looking for another possible answer based on the dynamic motions
of the IRDCs and the electric currents created by the charged
particles. Despite the low ionization fraction in an IRDC, can the
dynamic motions of these charged particles amplify the magnetic
field strengths within it? In this research we try to answer this
question.

In this paper, we focus our attention on the dynamic motions and
magnetic fields of the G34, of which we currently have suitable
observational information. We want to know if the dynamic motions of
this typical IRDC can increase the large-scale magnetic field
strengths within it? For this purpose, in section~2, we extract the
dynamic motions of the G34 from observational data and investigate
its rotation. We attribute a large-scale current density to this
rotating cloud by considering the velocity differences between the
positive and negative particles with very low ionization fractions.
The magnetic field resulting from this current density is obtained
in \S~3 and is compared with the observational results. Finally,
\S~4 is dedicated to summary and conclusion.

\section{Rotation and Current Density}

\begin{figure}
\epsscale{0.6} \center \plotone{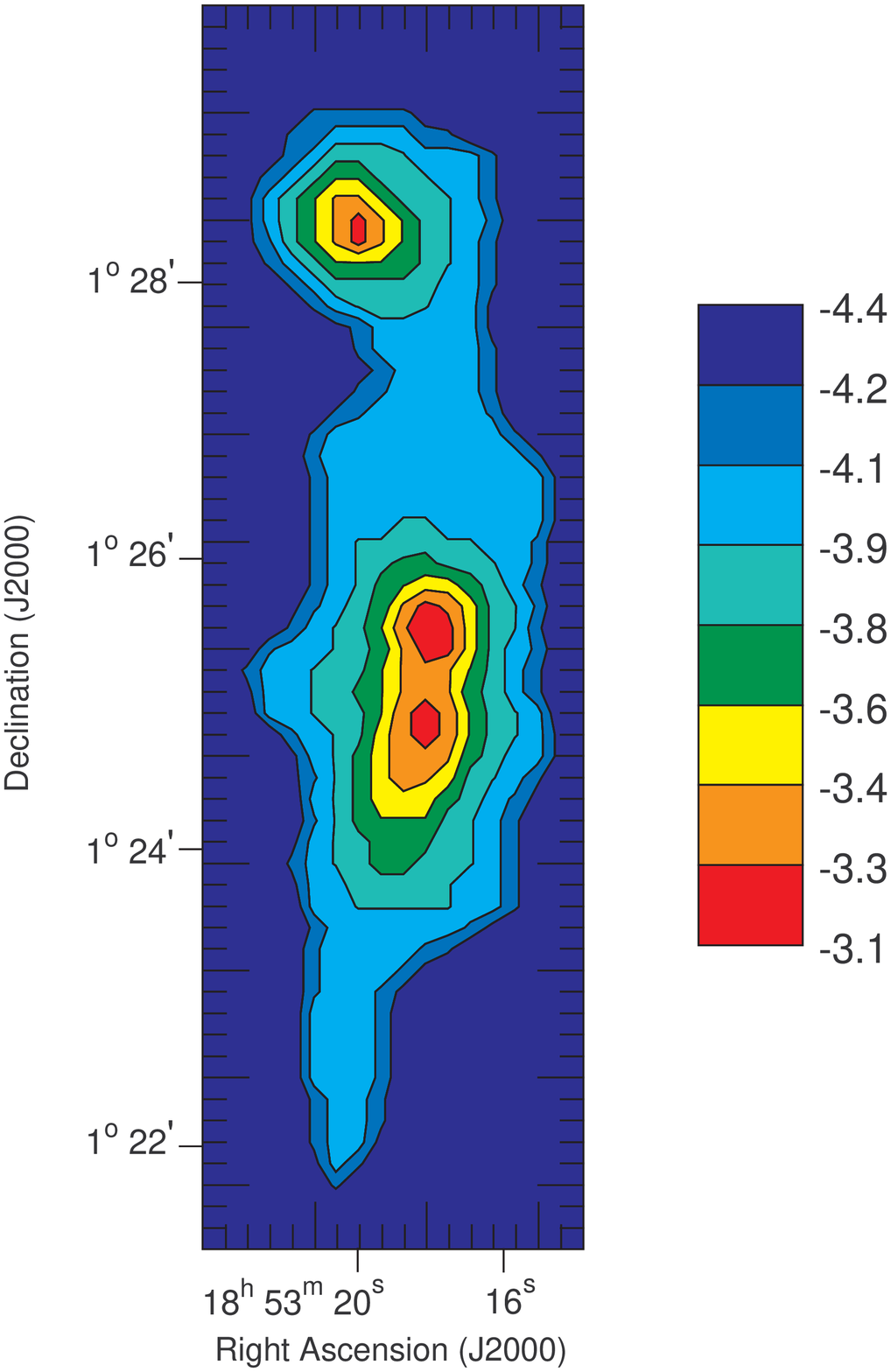}\\
\caption{The contour color-fill for the logarithm of mean number
density of ions, $\bar{n}_i$, in the unit of $\mathrm{cm}^{- 3}$.
The contours represent the isodensity points on the plane-of-sky and
the faraway contour represents the periphery of the
G34.\label{denl}}
\end{figure}

The molecular clouds are affected by various factors for ionization.
One of these ionizing factors is the cosmic rays, that in a steady
ionization equilibrium, lead to $\bar{n}_i\sim 7.5 \times 10^{-3}
(n_{n}/ 10^{3} \mathrm{cm}^{-3})^{1/2}$, where $\bar{n}_i$ and
$n_{n}$ are the mean number densities of ions and neutral particles
in the unit of $\mathrm{cm}^{-3}$, respectively (e.g., Shu~1992,
p.~362). Priestley, Wurster \& Viti~(2019) have recently
investigated differences in the molecular abundances and found that
the mean number density of ions follows an approximate power-law
relation with the number density of neutral particles as
\begin{equation}\label{denfrac}
    \bar{n}_i\sim 10^{-4} \left( \frac{n_{n}}{10^{3}
    \mathrm{cm}^{-3}}\right)^{0.4}\, \mathrm{cm}^{-3}.
\end{equation}
One of the remarkable interstellar molecular ion is
$\mathrm{N}_2\mathrm{H}^+$, which is widely used as suitable tracer
molecules in the study of molecular clouds (Thaddeus \&
Turner~1975). The abundance and distribution of
$\mathrm{N}_2\mathrm{H}^+$ is well explained by the ion-molecule
chemistry (Womack, Ziurys \& Wyckoff~1992). The integrated emission
of the $\mathrm{N}_2\mathrm{H}^+$ line can provide suitable
situations for determining the $\mathrm{H}_2$ density in the
molecular clouds and vice versa. Here, we use the integrated
emission of the $\mathrm{N}_2\mathrm{H}^+$ line detected by Tang et
al.~(2019, Figure~3) for the IRDC G34. The
$\mathrm{N}_2\mathrm{H}^+$ line intensity is actually proportional
to the column density along the line-of-sight. We assume that G34
has a uniform slab geometry in the line-of-sight direction so that
the column density implicitly represents the volume density. Womack,
Ziurys \& Wyckoff~(1992) concluded that the fractional abundances of
$\mathrm{N}_2\mathrm{H}^+$, relative to H2, is $\sim 4\times
10^{-10}$ toward both the warm and cold clouds. Thus, we assume that
the fractional abundance of this ion, relative to the hydrogen
molecules, is constant across the G34. We divide Figure~3 of Tang et
al.~(2019) into $17 \times 58$ square pixels with side $\sim 0.16 \,
\mathrm{pc}$. Sanhueza et al.~(2010) worked on the clumps in the G34
region and concluded that their number densities are between
$1.8\times 10^4\, \mathrm{cm}^{-3}$ to $3.9\times 10^5\,
\mathrm{cm}^{-3}$. On the other hand, in general, the number density
around the molecular clouds drops to $10^2\, \mathrm{cm}^{-3}$
(e.g., Ballesteros-Paredes et al.~2020). Therefore, here, we assume
that the $\mathrm{H}_2$ density in the centers of the clumps MM1,
MM2, and MM3 of G34, is in the order of $10^5\, \mathrm{cm}^{-3}$,
and in the outer and more distant periphery of G34, the number
density of neutral particles drops to $10^2\, \mathrm{cm}^{-3}$. In
this way, we consider the outermost contour in the Figure~3 of Tang
et al.~(2019) as the periphery of G34 and assign a density of
$10^2\, \mathrm{cm}^{-3}$ to this contour. We also assign the
innermost contour to the density of $10^5\, \mathrm{cm}^{-3}$.
Assuming a simple linear relationship between the integrated
emission of the $\mathrm{N}_2\mathrm{H}^+$ line and the column
density, we can determine the number density of neutral particles
assigned to the other contours. Thus, according to the number
density of neutral particles obtained from the integrated emission
of the $\mathrm{N}_2\mathrm{H}^+$ line, and the correlation relation
between ion and neutral particle densities as outlined by the
equation (\ref{denfrac}), we can determine the mean number density
of ions inside G34. The result is shown in Fig.~\ref{denl} as a
contour color-fill map.

The right ascension and declination of the center of G34 is
approximately equal to $RA = 18\,\mathrm{h}\, 53\,\mathrm{m}\,
20\,\mathrm{s} = 283.3^\circ$ and $DEC = + 1^\circ\, 26' =
1.43^\circ$ , which in the galactic coordinates is equal to $l = +
34.42^\circ$ and $b = + 0.23^\circ $. Stars and interstellar matters
closer to the Galactic center complete their orbits in less time
than those of the others further out. According to the Galactic
longitude of G34, we can deduce that it moves away from us so that
the measured line-of-sight centroid velocities present the red-shift
motion of the cloud gas (e.g., Sparke \& Gallagher~2007). Tang et
al.~(2019) extracted the kinematic information of the G34 using the
$\mathrm{N}_2\mathrm{H}^+ \; (J=1-0)$ line. Their centroid velocity
map is very robust, with a typical velocity uncertainty of $\simeq
0.1\, \mathrm{km\,s}^{-1}$. They reported the maps of the centroid
velocity and the dispersion of the $\mathrm{N}_2\mathrm{H}^+$ line
in Figure~3. The results show that the G34 presents a very organized
velocity field throughout the filament. The velocity gradient is
along with the East-West direction, consistent with measurements in
$\mathrm{N}\mathrm{H}_3$ at a higher angular resolution of $3''$ by
Dirienzo et al.~(2015).

\begin{figure}
\epsscale{0.7} \center \plotone{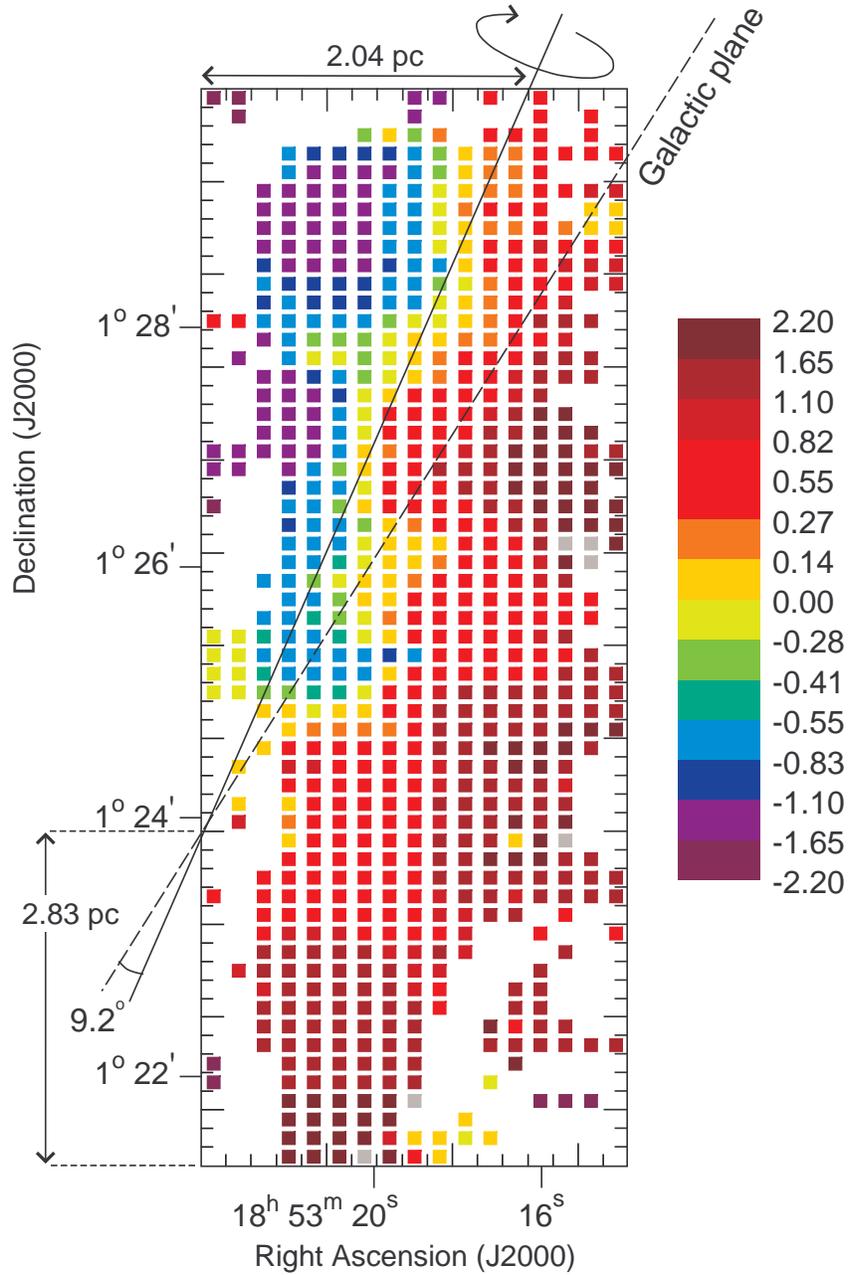}\\
\caption{The color-pixel-map of the line-of-sight centroid velocity,
relative to the median, in the unit of
$\mathrm{km}\,\mathrm{s}^{-1}$. Pixels with a relative velocity of
approximately zero (median value) are shown with a solid
hypothetical line, which represents the axis of rotation, and has an
approximate angle of $9.2^\circ$  to the Galactic plane. We do not
have enough information about the line-of-sight centroid velocity at
the boundaries of  G34, and therefore these pixels are shown in
white color.\label{vel}}
\end{figure}

If we assume that the filamentary structure of G34 is created by the
shock compression of a turbulent inhomogeneous molecular cloud
(e.g., Inoue et al.~2018), the velocity gradient perpendicular to
the filament axis can be considered as tracing the matter flow onto
a sheet-like structure (e.g., Arzoumanian et al.~2018). Here we
consider another possible choice in which the velocity gradient is
related to the rotation. If we assume that the materials of G34 were
rotating as a bulk body with its rotation axis perpendicular to the
line-of-sight, the measured velocity gradient would correspond to
$\Omega$, the cloud's angular velocity. A velocity gradient of
magnitude $\sim 1.7 \,\mathrm{km} \,\mathrm{s}^{-1}
\,\mathrm{pc}^{-1}$, from left to right in Figure~3 of Tang et
al.~(2019), can be seen. This value implies that G34 rotates slowly
with a representative period of $2\pi /\Omega \sim 3.5 \times 10^6
\, \mathrm{yr}$. We divide the right ascension direction in Figure 3
of Tang et al.~(2019) into 17 pixels and the declination direction
into 58 pixels. We extract the line-of-sight centroid velocity for
each pixel and subtract it from their median value. These relative
velocities for each pixel (relative to the median value) are shown
in Fig.~\ref{vel} as color-pixel-map. As shown in the figure, a
hypothetical line can be passed through some pixels that have an
approximately zero relative velocity (the median value). This
assumptive line can represent the axis of rotation, which is shown
in Fig.~\ref{vel}, and is approximately parallel to the plane of
Galaxy (with an angle near to $9.2^\circ$). Thus, it can be
visualized that the G34 rolls around the plane of the Galaxy like a
rolling cylinder with a slow angular velocity $\Omega\sim 5.7\times
10^{-14}\,\mathrm{s}^{-1}$, and this rolling represents part of the
Galaxy's rotation around its center.

The moving charged particles with density $n$, electric charge $q$,
and velocity $\mathbf{v}$ can produce a current density as
$\mathbf{J}=nq\mathbf{v}$. The true current density in IRDCs is
carried by a minor fraction of charged species: free electrons,
ions, and grains. Since the number density of grains is so low that
they contribute negligibly to the current, we have
\begin{equation}\label{curie}
    \mathbf{J} = \sum_i n_i q_i \mathbf{v}_i - n_e e \mathbf{v}_e,
\end{equation}
where $n_i$ is the number density of each species of ions with
charge $q_i$ and velocity $\mathbf{v}_i$. The number density of free
electrons, with charge $e$ and velocity $\mathbf{v}_e$, is denoted
as $n_e$. If we demand an ideal complete charge neutrality through
the G34, we have $n_e \approx \sum_i n_i (q_i/e)$. But, in the real
world, the charge neutrality is not complete. Also, there is not
enough information about the velocities of electrons and various
ions in G34. Here, we assume that the ions are well coupled with the
neutral matter of the cloud and move with velocity $\mathbf{v}$. In
this model, the free electrons can move forward or backward more
quickly. In this way, according to the lack of information, we
approximately apply the relation $J_{\perp}\approx \zeta e \bar{n}_i
v_{\perp}$ for the current density component along the
line-of-sight, where $\zeta$ has a positive and/or negative value
depending on the non-completeness of charge neutrality and the
relative speed of ions and electrons at each location of the cloud,
and $v_{\perp}$ is the velocity along the line-of-sight (such as one
which is deduced from $\mathrm{N}_2 \mathrm{H}^+$ molecules in G34,
and is depicted in Fig.~\ref{vel}).

\begin{figure}
\epsscale{0.7} \center \plotone{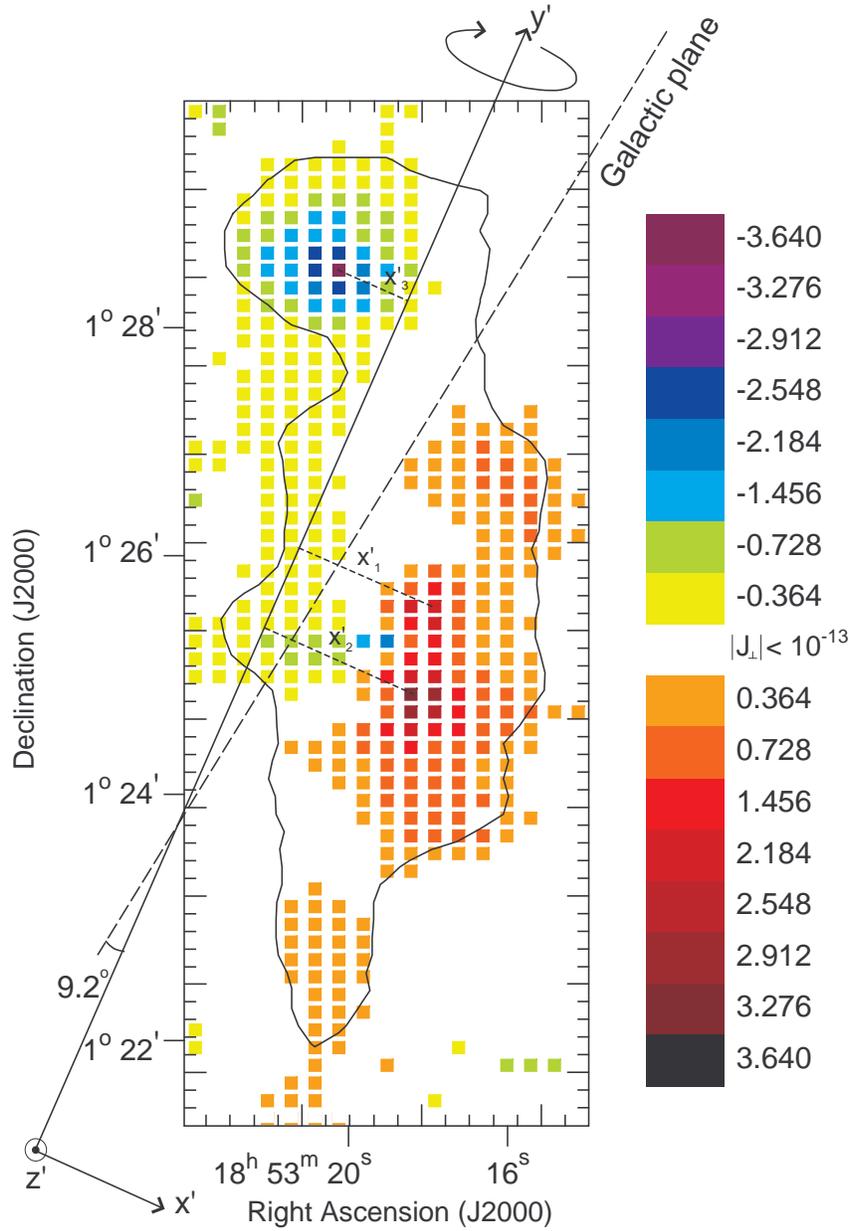}\\
\caption{The color-pixel-map of the component of current density
along the line-of-sight in the unit of $ 10^{-12}\,
\mathrm{esu}\,\mathrm{s}^{-1}\, \mathrm{cm}^{-2}$. The contour
represents the periphery of G34 and the outer pixels are not
significant. The Galactic plane is shown by the dashed line. We
choose the $x'y'z'$ coordinate system so that the $z'$-axis is in
the line-of-sight and the $y'$-axis coincides on the rotation axis,
which has angle $9.2^\circ$ concerning the Galactic plane. The
distances from the center of the clumps MM1, MM2, and MM3 to the
rotation axis are indicated by $x'_1$, $x'_2$, and $x'_3$,
respectively. \label{cur}}
\end{figure}

To our knowledge, there is not as much research on the electric
currents in the interstellar medium, but we can refer to the article
Carlqvist \& Gahm~(1992), who draw attention to the electric
currents in the sub-filaments in the molecular clouds. This study
shows that axial currents of the order of a few times
$10^{13}\,\mathrm{A}$ are necessary for the clouds to be in
equilibrium. We assume that the same order of magnitude of electric
current is also in the G34. In this way, if we consider the electric
current of each pixel in the G34 to be in the order of a few times
$10^{13}\,\mathrm{A}$, then the current density component along the
line-of-sight of each square pixel with side $\sim
0.16\,\mathrm{pc}$ will be of the order of a few times $10^{-12}\,
\mathrm{esu}\,\mathrm{s}^{-1}\, \mathrm{cm}^{-2}$. In this way,
using typical values in the clumps of G34, the current density along
the line-of-sight is
\begin{equation}\label{Jperp}
    J_{\perp}= 10^{-12} \left( \frac{\zeta}{10^{-4}} \right)
    \left( \frac{\bar{n}_i}{5 \times 10^{-4}\,\mathrm{cm}^{-3}} \right)
    \left( \frac{v_\perp}{1\,\mathrm{km}\,\mathrm{s}^{-1}} \right)
    \, \frac{\mathrm{esu}}{\mathrm{cm}^2\mathrm{s}}.
\end{equation}
The incompleteness of charge neutrality and the relative velocity of
free electrons to the ions and neutral particles in the molecular
clouds are not available in the observational data yet. Therefore,
identifying the precise value of $\zeta$, in the various locations
of molecular clouds, requires the challenge of theoretical research
and suitable simulations. For example, the relative ion-electron
drift in the non-ideal magnetohydrodynamics and the degree of charge
neutrality can be extracted from some numerical codes such as NICIL
(Wurster~2016, Wurster~2021). In this way, the value of the
parameter $\zeta$ may be estimated as a function of the gas
properties, but this subject is out of the scope of this paper.
Since we do not currently have a suitable function for $\zeta$, we
choose a constant value as a simple step in this research. Finally,
what encourages us for this choice of $\zeta$ is that the magnetic
field strengths inside the G34, resulting from the choice of the
current density of the order of a few times $10^{-12}\,
\mathrm{esu}\,\mathrm{s}^{-1}\, \mathrm{cm}^{-2}$, consistently
corresponds to the observational results. Thus, by choosing a
constant value for $\zeta$, knowing the mean number density of ions
as depicted in the Fig.~\ref{denl}, and given the relative
velocities extracted for each pixel of the G34 as shown in the
Fig.~\ref{vel}, we can obtain an approximate value for the mean
current density component along the line-of-sight. The results of
this current density component inside the G34, with a typical value
of $\zeta\approx +10^{-4}$, are shown in Fig.~\ref{cur} as a
color-pixel-map.

\begin{figure}
\epsscale{0.7} \center \plotone{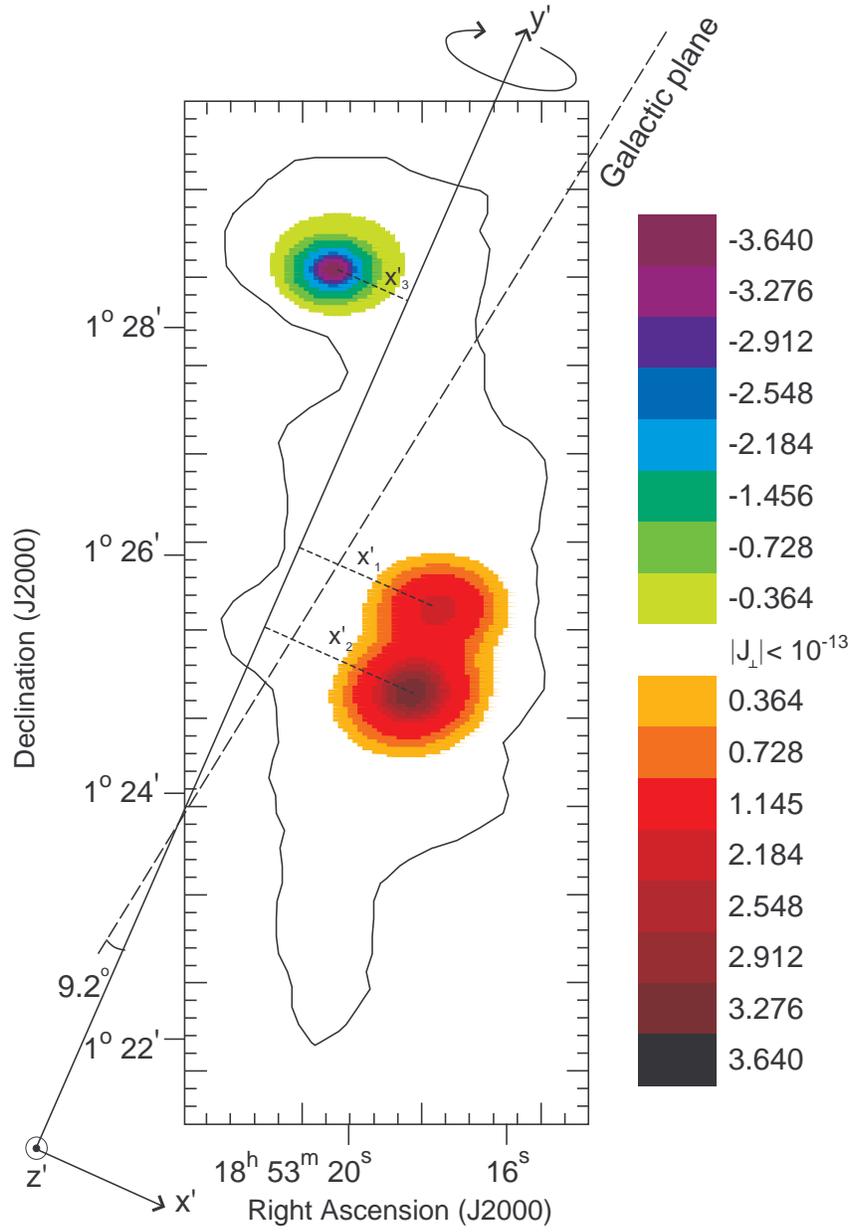}\\
\caption{The same as Fig.~\ref{cur} but with the mathematical model
(\ref{currentG}) and more pixels ($85 \times 290$). \label{curg}}
\end{figure}

As can be seen in Fig.~\ref{cur}, the current density in the center
of each clump is maximum, and it decreases as we move away from the
clump center. Therefore, it seems that a suitable mathematical model
can be formed to describe the large-scale current density in G34.
Here, we discard the useless data and apply this mathematical model
to investigate the magnetic field inside the G34 theoretically. A
suitable mathematical model for the current density in the
line-of-sight that corresponds to the results of Fig.~\ref{cur}, is
the sum of three normal Gaussian functions,
\begin{equation}\label{currentG}
    J_{z'}(x',y')= J_1\,
    e^{{-\left(\frac{|\mathbf{r'}-\mathbf{r'}_1|}{R_1}\right)^2}}
    +J_2\,
    e^{{-\left(\frac{|\mathbf{r'}-\mathbf{r'}_2|}{R_2}\right)^2}}
    -J_3\,
    e^{{-\left(\frac{|\mathbf{r'}-\mathbf{r'}_3|}{R_3}\right)^2}},
\end{equation}
where $R_1=0.34\,\mathrm{pc}$, $R_2=0.38\,\mathrm{pc}$, and
$R_3=0.21\,\mathrm{pc}$ are effective radius of clumps MM1, MM2, and
MM3, respectively, and $J_1\approx 2.4 \times 10^{-12} \,
\mathrm{esu}\,\mathrm{s}^{-1}\, \mathrm{cm}^{-2}$, and $J_2\approx
J_3\approx 3.6 \times 10^{-12} \, \mathrm{esu}\,\mathrm{s}^{-1}\,
\mathrm{cm}^{-2}$ are the maximum current density at the center of
clumps: $\mathbf{r'}_1=1.03\, \hat{i'} + 5.23\, \hat{j'}$,
$\mathbf{r'}_2=1.10\, \hat{i'} + 4.46\, \hat{j'}$, and
$\mathbf{r'}_3=-0.52\, \hat{i'} + 7.83\, \hat{j'}$ (all in the unit
of $\mathrm{pc}$), respectively. Using this mathematical model helps
us increase the number of pixels and to increase the accuracy of
theoretical calculations. The results of the mathematical model
(\ref{currentG}), with the discarded non-significant observational
data in the outer parts of the periphery of G34 and neglected values
at $|J_\perp| < 10^{-13} \, \mathrm{esu}\,\mathrm{s}^{-1}\,
\mathrm{cm}^{-2}$, are plotted in Fig.~\ref{curg}, which is exactly
equivalent to the Fig.~\ref{cur} but with more accurate pixels
(e.g., here $85 \times 290$).

\section{Magnetic Field Morphology}

The observations show that G34 is immersed inside an external
magnetic field approximately parallel to the Galactic plane (e.g.,
Jones et al.~2016). The Galactic magnetic fields are of the order of
some micro-Gauss (e.g., Jansson \& Farrar~2012), while the
observations show that the magnetic field strengths in G34 are in
the order of a few hundred $\mu \mathrm{G}$. (e.g., Soam et
al.~2019). If we want to give a theoretical explanation for this
huge increase of the magnetic field strengths, it seems that the
bulk motions of the charged particles in the G34 are effective.
Therefore, we can deduce that the current density, resulting from
the bulk motion of the charged particles within G34, can strengthen
the magnetic field and change its configuration.

\begin{figure}
\epsscale{0.7} \center \plotone{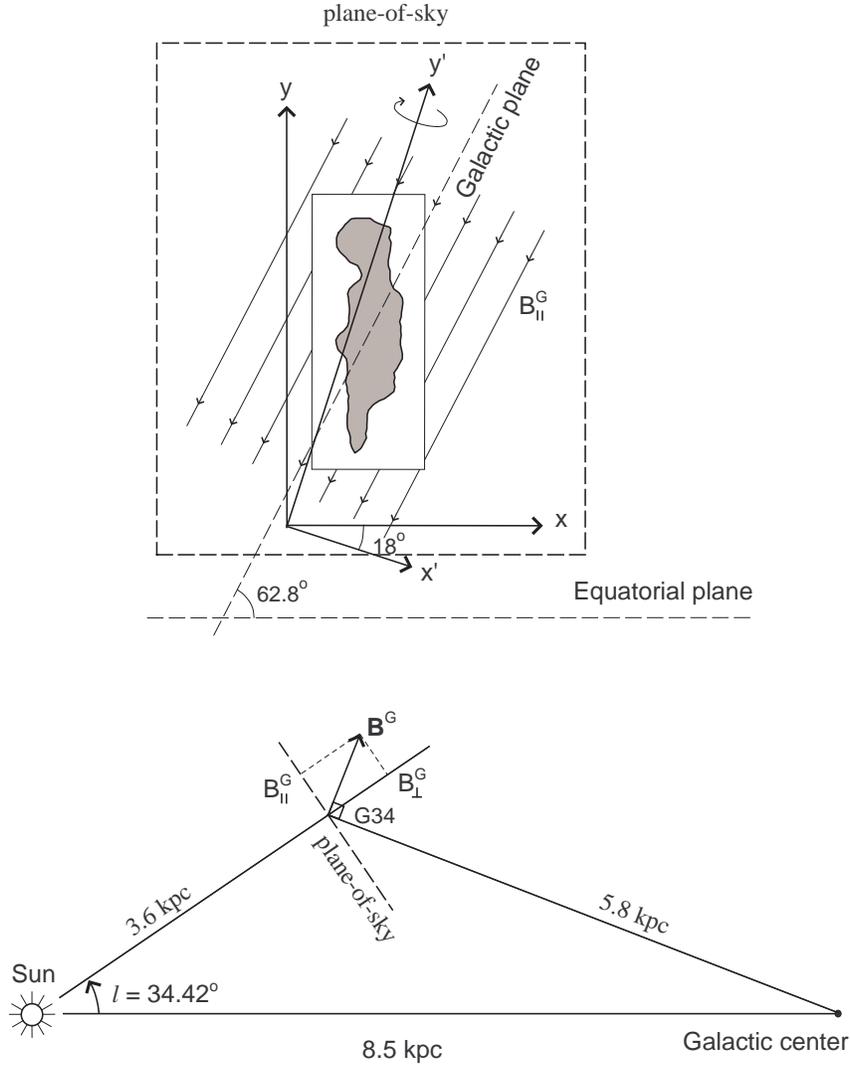}\\
\caption{Lower panel: the location of IRDC G34 on the Galactic plane
and the components of the Galactic magnetic field in this region.
Upper panel: the apparent view of G34 on the plane-of-sky. In the
upper panel, the equatorial and Galactic planes are shown as dashed
lines, and the Galactic magnetic field component on the plane-of-sky
around the G34 is shown uniformly and parallel to the Galactic
plane. The primed coordinate system is chosen so that $y'$
corresponds to the rotation axis and $z'$ is in the line-of-sight.
The $xyz$ coordinate system is obtained from rotation of the
$x'y'z'$ coordinate around the $z'$-axis so that the $x$-axis is
parallel to the equatorial plane. \label{schem}}
\end{figure}

The IRDC G34 is located on the Galactic plane, approximately
$5.8\,\mathrm{kpc}$ from the Galactic center (as depicted
schematically in Fig.~\ref{schem}). The studies of background
starlight polarimetry, dust emission polarimetry, Faraday rotation,
and synchrotron emission indicate that the large-scale Galactic
magnetic field is approximately parallel to the Galactic plane and
perpendicular to Galactocentric radius (e.g., Heiles~1996, Pshirkov
et al.~2011, Jansson \& Farrar~2012, Planck Collaboration et
al.~2015, Zenko et al.~2020). In this research, the external
magnetic field at the location of G34 is assumed to be a constant
value of $3\,\mu G$ and its direction is parallel to the Galaxy
plane and perpendicular to the Galactocentric radius. This magnetic
field, $\mathbf{B}^G$, has two components: $B^G_\parallel$ in the
plane-of-sky, and $B^G_\perp$ in the line-of-sight.

While we do not have enough information on the $B_\perp$, there is
appropriate observational information for $B_\parallel$ within the
G34. For example, Tang et al.~(2019) investigated the magnetic field
morphology across the G34 and its clumps MM1, MM2, and MM3 using the
polarization of thermal dust at a wavelength of $350\,\mu
\mathrm{m}$ with an angular resolution of $10''$
($0.18\,\mathrm{pc}$). Soam et al.~(2019) also presented the
magnetic fields mapped in IRDC G34 using $850\,\mu \mathrm{m}$ at
the James Clerk Maxwell telescope. They obtained a plane-of-sky
magnetic field strength, $B_\parallel$, of $470\pm 190\,\mu
\mathrm{G}$ in the central part of the cloud (near MM1/MM2), $100\pm
40\,\mu \mathrm{G}$ in the northern part of the cloud (near MM3) and
$60\pm 34\,\mu \mathrm{G}$ in the southern part.

Theoretically, we expect that the current density (\ref{currentG}),
which is generated by the cloud rotation, amplifies the magnetic
field strengths and changes its structure inside the G34. According
to Ampere's law,
\begin{equation}\label{Ampere}
    \nabla\times \mathbf{B}= \frac{4\pi}{c} \mathbf{J},
\end{equation}
and using the magnetic vector potential, $\mathbf{A}$, with the
Coulomb gauge, $\nabla\cdot \mathbf{A}=0$, we have
\begin{equation}\label{AmpA}
      \nabla^2 \mathbf{A}=- \frac{4\pi}{c} \mathbf{J}.
\end{equation}
We assume that G34 rotates with azimuthal symmetry around the
$y'$-axis shown in Fig~\ref{schem}. In this way, there is no
$y'$-component of the current density. If we average its
$x'$-component in the line-of-sight, $z'$, we approximately have
\begin{equation}\label{mjx}
     \frac{1}{\Delta z'}\int J_{x'} dz'\approx 0,
\end{equation}
where $\Delta z$ is the length along the line-of-sight, and the
azimuthal symmetry around the $y'$-axis is used. Therefore, the
Ampere's law (\ref{AmpA}) in the $x'y'z'$ coordinates can be
rewritten as
\begin{equation}\label{Ampxyp}
    \nabla^2 \bar{A}_{x'}= \nabla^2 \bar{A}_{y'}=0,
\end{equation}
and
\begin{equation}\label{Ampzp}
    \nabla^2 \bar{A}_{z'}= -\frac{4\pi}{c} \bar{J}_{z'},
\end{equation}
where the components of the magnetic vector potential are averaged
in the line-of-sight and $\bar{J}_{z'}$ is given approximately by
the sum of three normal Gaussian functions (\ref{currentG}).

With the rotation of the primed coordinate system through an angle
$18^\circ$ about the $z'$ axis, the equations (\ref{Ampxyp}) and
(\ref{Ampzp}) convert to the following equations in the $xyz$
coordinate system
\begin{equation}\label{Ampxy}
    \nabla^2 \bar{A}_{x}= \nabla^2 \bar{A}_{y}=0,
\end{equation}
\begin{equation}\label{Ampz}
    \nabla^2 \bar{A}_{z}(x,y)= -\frac{4\pi}{c} \bar{J}_{z}(x,y),
\end{equation}
where
\begin{equation}\label{coorx}
    x=x' \cos 18^\circ + y' \sin 18^\circ,
\end{equation}
\begin{equation}\label{coory}
    y=-x' \sin 18^\circ + y' \cos 18^\circ.
\end{equation}

We consider a uniform Galactic magnetic field $B^G\approx 3\,\mu
\mathrm{G}$ in the region of G34 (Jansson \& Farrar~2012). The
magnetic field components in the $xyz$ coordinate system are
\begin{equation}\label{BGx}
    B_x^G=-B^G\sin 34.5^\circ \sin 27.2^\circ,
\end{equation}
\begin{equation}\label{BGy}
    B_y^G=-B^G\sin 34.5^\circ \cos 27.2^\circ,
\end{equation}
\begin{equation}\label{BGz}
    B_z^G=-B^G\cos 34.5^\circ.
\end{equation}
The line-of-sight component of the magnetic field is $B^G_\perp
=-B^G_z$ and its component on the plane-of-sky is $B^G_\parallel=
\sqrt{{B^G_x}^2+{B^G_y}^2}$ as is depicted in Fig.~\ref{schem}. The
vector potential of a uniform magnetic field can be expressed in
different ways. Here, we use the following symmetric form
\begin{equation}\label{AGx}
    A^G_x=-\frac {1}{2} B^G_z y,
\end{equation}
\begin{equation}\label{AGy}
    A^G_y=\frac {1}{2} B^G_z x,
\end{equation}
\begin{equation}\label{AGz}
    A^G_z=B^G_x y - B^G_y x.
\end{equation}

The components of the averaged magnetic vector potential in the
$xyz$ coordinate system are represented by the equations
(\ref{Ampxy}) and (\ref{Ampz}), and must also satisfy the boundary
conditions (\ref{AGx})-(\ref{AGz}) for the periphery of G34. The
$\bar{A}_x$ and $\bar{A}_y$ components satisfy the Laplace equation,
(\ref{Ampxy}), with Dirichlet boundary conditions (\ref{AGx}) and
(\ref{AGy}). It is clear that there are unique and well-behaved
solutions for these components inside the bounded region as $
\bar{A}_x=\frac {1}{2} B^G_\perp y$ and $\bar{A}_y=-\frac {1}{2}
B^G_\perp x$. The $\bar{A}_z$ component of the magnetic vector
potential satisfies the Poisson equation in the 2-D Cartesian
coordinates, (\ref{Ampz}), with the boundary condition (\ref{AGz}).

We want to find a single static function $\bar{A}_z(x,y)$ which
satisfies the equation (\ref{Ampz}) within the region of G34, and
the desired boundary condition (\ref{AGz}). This problem cannot be
solved analytically, and thus some numerical methods must be used.
Since this is a boundary value (static) problem, the goal of a
numerical method is somehow to converge on the correct solution
everywhere at once. There are many methods to find an approximate
numerical solution for the Poisson equation. For example, the
equation can be discretized, and it can then be solved by relaxation
methods and/or rapid methods such as Fourier and cyclic reduction
methods (e.g., Press et al.~1992).

\begin{figure}
\epsscale{0.9} \center \plotone{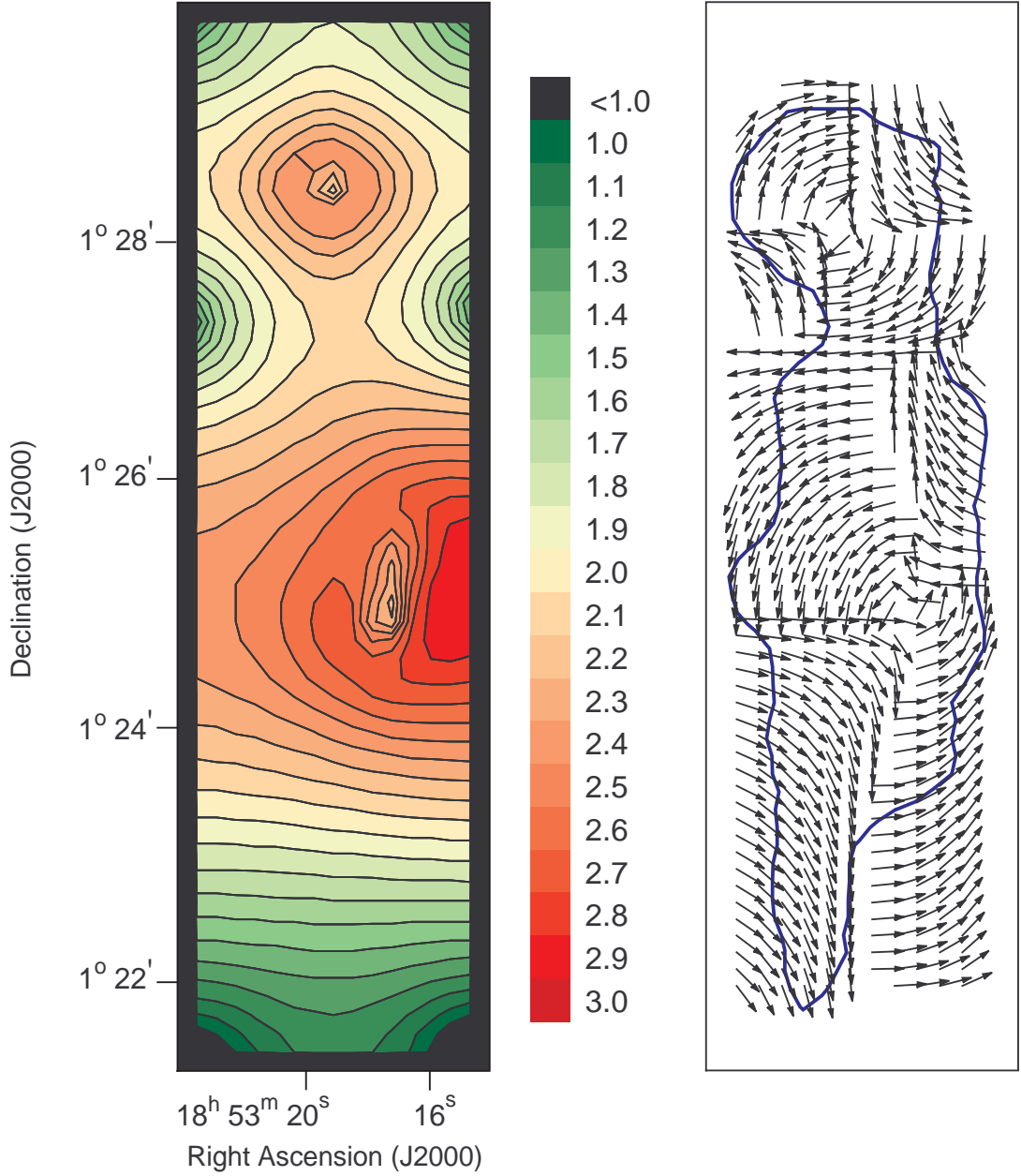}\\
\caption{The contour color-fill of the logarithm of magnetic field
strengths of G34 on the plane-of-sky in the unit of $\mu\mathrm{G}$
(left panel), and magnetic field orientations which are depicted by
unit vectors (right panel). \label{magl}}
\end{figure}

Because all the conditions on the boundary must be satisfied
simultaneously, the problem reduces to the solution of large numbers
of simultaneous algebraic equations. Many implementations of Poisson
solvers exist, but their software implementation is often not
publicly available, or it is a part of a larger software package.
One of the more general pioneer ones is the package FISHPACK, which
solves the second-order finite difference approximation of the
Poisson or Helmholtz equation on rectangular grids (Adams,
Swarztrauber \& Sweet~2016). Here, we use the FISHPACK to solve the
Poisson equation (\ref{Ampz}) with boundary condition (\ref{AGz}) in
2-D rectangular Cartesian coordinates: $0.6\leq x\leq 3.3 $ and
$0.1\leq y\leq 9.2 $ all in the unit of $\mathrm{pc}$. By
determining the magnetic vector potential for the pixels within the
G34, the components of the magnetic field are obtained via
$B_x=\partial \bar{A}_z(x,y)/\partial y$ and $B_y=-\partial
\bar{A}_z(x,y)/\partial x$. In this way, the strength and direction
of the magnetic field component on the plane-of-sky can also be
determined. The magnetic field strength, $B_\parallel= \sqrt{B_x^2 +
B_y^2}$, and its orientation on the plane-of-sky are shown in
Fig.~\ref{magl}.

\begin{figure}
\epsscale{0.55} \center \plotone{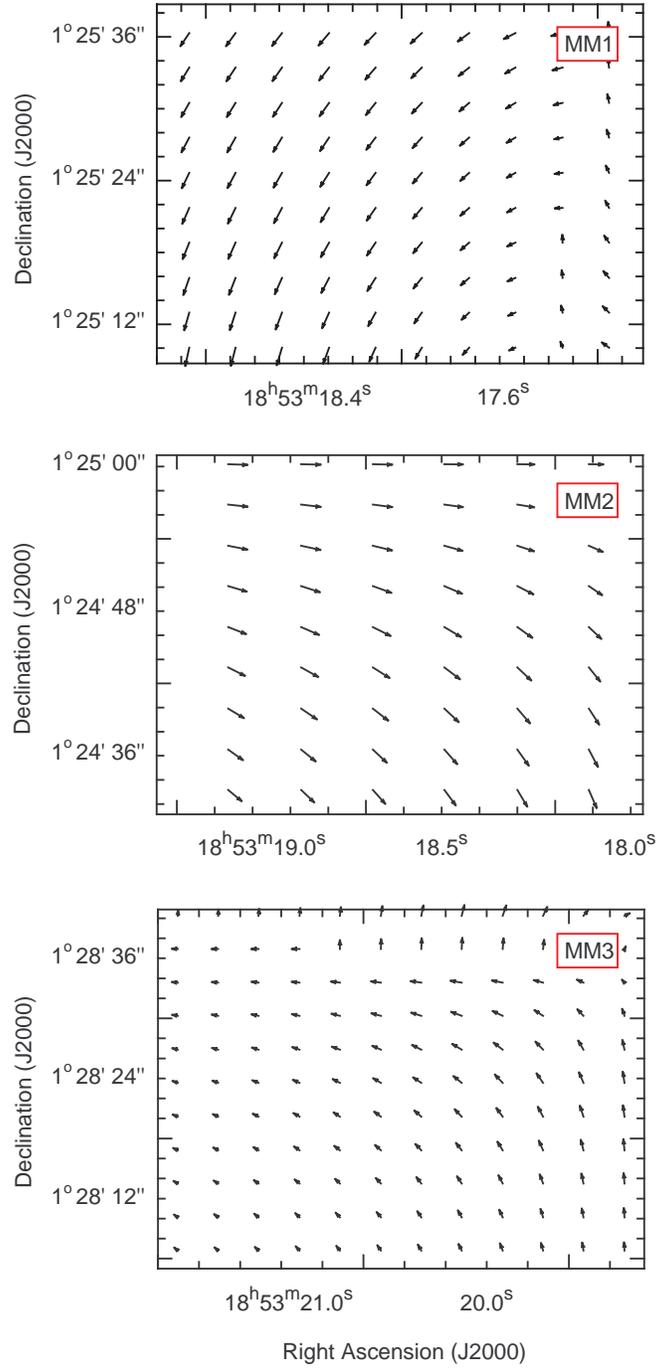}\\
\caption{The vector-plot of plane-of-sky magnetic fields of clumps
MM1, MM2 and MM3.\label{magm}}
\end{figure}

The panels (d)-(f) in Figure~2 of Tang et al.~(2019) display dust
continuum (gray scale) and $B$-field segments (blue segments)
observed with the higher angular resolution for MM3, MM1 (Hull et
al.~2014), and MM2 (Zhang et al.~2014). Also, the red segments show
the $B$-field detected with SHARP, which has presented the
observational fact that a mostly uniform large-scale $B$-field
perpendicular to the filament is observed toward MM1 and MM2, while
a bending $B$-field is seen closely aligned with the MM3 major axis.
Thus, it is interesting to look at the strengths and orientations of
our theoretical model for magnetic field morphology inside these
clumps, and compare them to the observational results. For this
purpose, a vector-plot of the magnetic field $B_\parallel$ toward
the clumps MM1, MM2, and MM3 is shown in Fig.~\ref{magm}.

\begin{figure}
\epsscale{0.8} \center \plotone{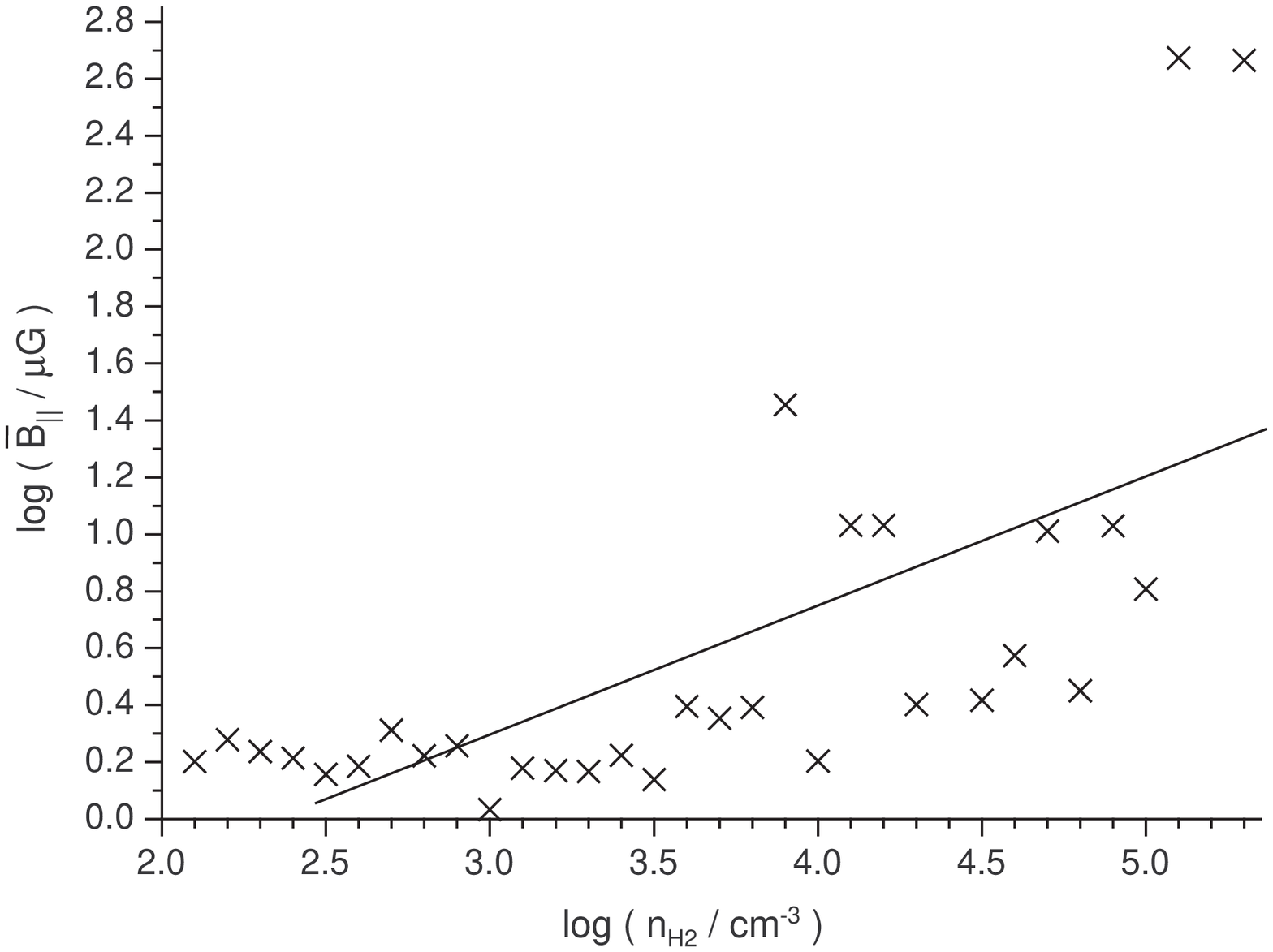}\\
\caption{The logarithmic plot of the mean magnetic field versus
$\mathrm{H}_2$ number density. The fitted straight-line indicates
the relation $\bar{B}_\parallel \propto n_{\mathrm{H}_2}^\eta $ with
power $\eta = 0.45 \pm 0.08$. \label{magd}}
\end{figure}

A good test for this theoretical model of magnetic field morphology
is scaling magnetic field strengths with density. This relation is
usually parameterized as a power-law, $B\propto \rho^\eta$, where
$\eta$ is a constant (Crutcher~2012). In the strong-field models,
which are applicable in most IRDCs, the density increases faster
than the magnetic field so that $\eta\lesssim 0.5$ is predicted
(e.g., Mouschovias \& Ciolek~1999). Given the map of mean number
density of ions in Fig.~\ref{denl} and the relative abundances of
ions to $\mathrm{H}_2$ by equation (\ref{denfrac}), we can estimate
the number density of $n_{\mathrm{H}_2}$ for each pixel within G34.
On the other hand, according to the left panel of Fig.~\ref{magl},
we can determine the strength of the magnetic field corresponding to
each pixel. The logarithm of the $\mathrm{H}_2$ number density,
$\log (n_{\mathrm{H}_2}/\mathrm{cm}^{-3})$, is between $2$ and $5.3$
and we divide it into equal segments with interval $h=0.1$. If we
denote the number of pixels in the range
\begin{equation}\label{Nj}
    2+(j-1) h< \log (n_{H_2}/\mathrm{cm}^{-3}) \leq 2+ jh; \quad
    j=1,2,...,33,
\end{equation}
by $N_j$, we can express the average magnetic field associated with
this density as
\begin{equation}\label{BNj}
    \bar{B}_\parallel= \frac{1}{N_j} \sum_{i=1}^{N_j}
    B_\parallel^{(i)},
\end{equation}
where $B_\parallel^{(i)}$ is the magnetic field strength at pixel
$i$. The relation between the mean magnetic field (\ref{BNj}) and
the density of G34 is shown as a logarithmic plot in
Fig.~\ref{magd}. If we consider a suitable straight-line, which is
approximately fitted to the set of data points in this figure, its
slope will be $\eta= 0.45 \pm 0.08$.

\section{Summary and Conclusion}

With the development of observational techniques, we nowadays have
good information about the magnetic fields of IRDCs. We know that
these objects are interstellar molecular clouds with a strong
magnetic field of the order of several hundred $\mu \mathrm{G}$. One
of these IRDCs is G34, to which it has been paid attention by
observers to obtain suitable observational data. It is located on
the plane of the Milky Way, approximately $5.8\, \mathrm{kpc}$ from
the Galactic center. The large-scale Galactic magnetic field at the
region of this IRDC is of the order of $\mu \mathrm{G}$, while the
magnetic field inside this cloud is of the order of several hundred
$\mu\mathrm{G}$. The question that led us to this research is
whether the dynamical motion of an IRDC like G34 can amplify the
magnetic field strength from some $\mu \mathrm{G}$ outside the cloud
to a few tenths of $\mathrm{mG}$ inside it?

The observational data show that G34 is moving away from us and that
its component of apparent motion, in the line-of-sight direction
(i.e., toward the plane-of-sky), is not uniform for all cloud
regions. The northwest part of the cloud is moving away from us
faster than the southeast part. We attributed this velocity gradient
to the cloud rotation and obtained an approximate rotation axis for
G34 that is indicated in Fig.~\ref{vel}. To obtain this figure, we
used the observational data of Tang et al.~(2019), who used the
dynamics of $\mathrm{N}_2\mathrm{H}^+$ molecules. The molecular
hydrogen number density from the outer edge of a typical IRDC to the
center of its clumps is about $ \sim 10 ^ 2-10 ^ 5 \, \mathrm {cm} ^
{- 3} $. According to the relative abundance of the ions in the
molecular clouds (e.g., equation \ref{denfrac}) , the contour
color-fill of the mean number density of ions in the G34 is plotted
in Fig.~\ref{denl}.

Having known the density of charged particles and the speed of their
rotation, we obtained an approximate relation for the line-of-sight
component of the current density, which is shown in Fig.~\ref{cur}.
Of course, to extract this figure, we assumed that there is a
constant value for the incompleteness of charge neutrality and the
velocity differences between the positive and negative particles
with very low ionization fractions inside the G34. The appearance of
Fig.~\ref{cur} shows that the current densities in the center of the
clumps MM1, MM2, and MM3 are the highest values and decrease as we
move away from the center of each clump. With this description, the
relation (\ref{currentG}), which is the sum of three normal Gaussian
functions, is presented to express the current density around the
rotation axis of G34 (Fig.~\ref{curg}). Section 3 is represented to
measure the amount of magnetic field strength and morphology
generated by current density (\ref{currentG}) inside the G34.

We have first schematically depicted Fig.~\ref{schem}, which
describes the location of G34 on the Galactic plane and its visible
image on the plane-of-sky. The Galactic magnetic field and the
rotation axis of G34 are also shown in this figure. The $xyz$
coordinate system is chosen so that the axis $ x $ is parallel to
the equatorial plane and the axis $z$ is in the line-of-sight. We
then averaged Ampere's law across the line-of-sight, which led to
the two-dimensional Poisson equation (\ref{Ampz}). The periphery of
the G34 is considered as rectangular borders with $ 0.6 \le x \le
3.3$ and $ 0.1 \le y \le 9.2$ (all in the unit of $\mathrm{pc}$), as
shown schematically in Fig.~\ref{schem}. In this way, the Poisson
equation (\ref{Ampz}) is solved numerically, using the FISHPACK with
boundary condition (\ref{AGz}), to obtain the mean vector potential
$\bar{A}_z(x,y)$ for the pixels inside this rectangle. By finding
out the mean vector potential, we obtained the magnetic field
components along with the $x$ and $y$ coordinates. Then, the
magnetic field strength and its orientation in the plane-of-sky is
obtained. The results are shown in Fig.~\ref{magl}. It can be seen
that the magnetic field strengths in the G34 are of the order of
several hundred $\mu\mathrm{G}$. Thus, the idea that the dynamical
motion of the charged particles can amplify the magnetic fields
through the IRDCs is somewhat established. The vector-plot of
plane-of-sky magnetic fields of clumps MM1, MM2, and MM3 are also
presented in Fig.~\ref{magm}. The results obtained for the magnetic
field orientation in the plane-of-sky are almost consistent with the
observational results and are somewhat convincing.

Finally, we proceeded to calculate the relationship between magnetic
field strengths and densities in the G34. This relation is expressed
for typical molecular clouds as $B\propto  \rho ^ \eta$
(Crutcher~2012). To do this, we divided the G34 image on the
plane-of-sky into pixels and determined the magnetic field and
density of each pixel. Then we obtained the average field strength
in each density range and plotted its logarithm versus the density
logarithm in Fig.~\ref{magd}. We also fitted a straight line at
these points. The slope of this line is approximately $\eta \approx
0.45$. This result is consistent with this subject that in the
molecular clouds with strong magnetic fields such as IRDCs, the
power $\eta$ should be less than $0.5$. Therefore, we can conclude
from this research that the dynamical motion of IRDCs, and
especially their rotations, can amplify the magnetic field within
the cloud to several hundred $ \mu \mathrm {G} $.

\section*{Data Availability}
No new data were generated or analyzed in support of this research.

\section*{Acknowledgments}
We appreciate the the anonymous reviewer for his/her careful
reading, useful comments and suggested improvements.



\clearpage

\clearpage

\clearpage

\clearpage

\clearpage

\clearpage

\clearpage

\clearpage


\begin{thebibliography}{99}

\bibitem[]{ada16} Adams J.C., Swarztrauber P.N., Sweet R., 2016, ascl:1609.004

\bibitem[]{ane20} A\~{n}ez-L\'{o}pez N., 2020, A\&A, 644, 52

\bibitem[]{arz18} Arzoumanian D., Shimajiri Y., Inutsuka S., Inoue T., Tachihara
K., 2018, PASJ, 70, 96

\bibitem[]{bal20} Ballesteros-Paredes J. et al., 2020, Space Science
Reviews, 216, 76

\bibitem[]{beu17} Beuther H. et al., 2017, ApJ, 836, 199

\bibitem[]{car98} Carey S.J.,  Clark F.O., Egan M.P., Price S.D., Shipman R.F., Kuchar T.A., 1998, ApJ, 508, 721

\bibitem[]{car92} Carlqvist P., Gahm G.F., 1992, IEEE Transactions
on Plasma Science, 20, 867

\bibitem[]{cru12} Crutcher R.M., 2012, ARA\&A, 50, 29

\bibitem[]{dir15} Dirienzo W.J., Brogan C., Indebetouw R., Chandler C.J., Friesen R.K., Devine K.E., 2015, AJ, 150, 159

\bibitem[]{ega98} Egan M.P., Shipman R.F., Price S.D., Carey S.J., Clark F.O., Cohen M., 1998, ApJ, 494, 199

\bibitem[]{hei96} Heiles C., 1996, ApJ, 462, 316

\bibitem[]{hen19} Hennebelle P., Inutsuka S., 2019, FrASS, 6, 5

\bibitem[]{hoq17} Hoq S., Clemens D.P., Guzm\'{a}n A.E., Cashman L.R., 2017, ApJ, 836, 199

\bibitem[]{hul14} Hull C.L.H., 2014, ApJS, 213, 13

\bibitem[]{ino18} Inoue T., Hennebelle P., Fukui Y., Matsumoto T., Iwasaki K., Inutsuka
S., 2018, PASJ, 70, 53

\bibitem[]{ise21} Isequilla N.L., Ortega M.E., Areal M.B., Paron S., 2021, A\&A, 649, 139

\bibitem[]{jan12} Jansson R., Farrar G.R., 2012, ApJ, 757, 14

\bibitem[]{jon16} Jones T.J., Gordon M., Shenoy D., Gehrz R.D., Vaillancourt J.E., Krejny M., 2016, AJ,
151, 156

\bibitem[]{juv18} Juvela M. et al., 2018, A\&A, 620, 26

\bibitem[]{lis20} Liu H., Sanhueza P., Liu T., Zavagno A., Tang X., Wu Y., Zhang S., 2020, ApJ, 901, 31

\bibitem[]{liz20} Liu J., Zhang Q., Qiu K., Liu H.B., Pillai T., Girart J.M., Li Z.,
Wang K., 2020, ApJ, 895, 142

\bibitem[]{mou99} Mouschovias T.Ch., Ciolek G.E., 1999, in Lada C.J., Kylafis N.D., eds, The Origin of Stars and Planetary Systems. NATO ASI Series C, Kluwer Academic Publishers, p. 305

\bibitem[]{mes66} Mestel L., 1966, MNRAS, 133, 265

\bibitem[]{per96} Perault M. et al., 1996, A\&A, 315, 165

\bibitem[]{pla15} Planck Collaboration et al., 2015, A\&A, 576, 104

\bibitem[]{pre92} Press W.H., Teukolsky S.A., Vetterling W.T., Flannery B.P., 1992, Numerical recipes in FORTRAN. The art of scientific computing. Cambridge University Press

\bibitem[]{psh11} Pshirkov M.S., Tinyakov P.G., Kronberg P.P., Newton-McGee K.J., 2011, ApJ, 738, 192

\bibitem[]{rat05} Rathborne J.M., Jackson J.M., Chambers E.T., Simon R., Shipman R., Frieswijk W., 2005, ApJ, 630, 81

\bibitem[]{rat06} Rathborne J.M., Jackson J.M., Simon R., 2006, ApJ, 641, 389

\bibitem[]{san10} Sanhueza P., Garay G., Bronfman L., Mardones D., May J., Saito M., 2010, ApJ, 715, 18

\bibitem[]{san16} Santos F.P., Busquet G., Franco G.A.P., Girart J.M., Zhang Q., 2016, ApJ, 832, 186

\bibitem[]{shu92} Shu F.H., 1992, The Physics of Astrophysics: Gas
Dynamics, University Science Books

\bibitem[]{soa19} Soam A. et al., 2019, ApJ, 883, 95

\bibitem[]{spa07} Sparke L.S., Gallagher III J.S., 2007, Galaxies in the Universe: An Introduction. Second Edition. Cambridge University Press, Cambridge, UK

\bibitem[]{tan19} Tang Y., Koch P.M., Peretto N., Novak G., Duarte-Cabral A., Chapman N.L., Hsieh P., Yen H., 2019, ApJ, 878, 10

\bibitem[]{tha75} Thaddeus P., Turner B.E., 1975, ApJ, 201, 25

\bibitem[]{wom92} Womack M., Ziurys L.M., Wyckoff S., 1992, ApJ, 387, 417

\bibitem[]{wur16} Wurster J., 2016, PASA, 33, 41

\bibitem[]{wur16} Wurster J., 2021, MNRAS, 501, 5873

\bibitem[]{zen20} Zenko T., Nagata T., Kurita M., Kino M., Nishiyama S., Matsunaga N., Nakajima Y., 2020, PASJ, 72, 27

\bibitem[]{zha14} Zhang Q., 2014, ApJ, 792, 116

\end{thebibliography}
\end{document}